\documentstyle[aps,twocolumn]{revtex}

\begin{document}

\title{\bf Non dissipative decoherence of Rabi oscillations}
\author{Rodolfo Bonifacio$^{1}$, Stefano Olivares$^{1}$, 
Paolo Tombesi$^{2}$, and David Vitali$^{2}$}
\address{$^{1}$Dip. di Fisica, Universit\`a di Milano,
INFN and INFM, Sezione di Milano, via Celoria 16, 20133, Milano, Italy \\
$^{2}$Dip. di Matematica e Fisica and Unit\`a INFM, Universit\`a di
Camerino, via Madonna delle Carceri 62032, Camerino, Italy}

\date{\today}

\maketitle

\begin{abstract}
We present a simple theoretical description of two recent experiments 
where damping of Rabi oscillations, which cannot be attributed to 
dissipative decoherence, has been observed. This is obtained considering the
evolution time or the Hamiltonian as random variables and then averaging
the usual unitary evolution on a properly derived, model-independent,
probability distribution.
\end{abstract}


Even though decoherence is a very general phenomenon \cite{zur}, 
it is very difficult
to verify it experimentally because most often 
the physical nature of the environmental degrees of freedom responsible
for the decoherence process remains unknown. The only controlled
experimental verification of decoherence has been given by the experiment
of Ref.~\cite{prlha}, in which the progressive transformation of a 
linear superposition of two coherent states of a microwave cavity mode
into the corresponding statistical mixture has been monitored.
In this case, the environmental decoherence has been checked with
no fitting parameters because its physical origin, i.e., photon leakage 
out of the cavity, was easily recognizable and measurable.
In this case, it is even possible to {\em control} decoherence,
i.e., to considerably suppress its effects, for example by using
appropriately designed feedback schemes 
\cite{prlno}.

In some cases however, the mechanisms responsible for decoherence are not 
easily individuated and examples are provided by two recent
experiments which observed Rabi oscillations between
two circular states of a Rydberg atom in a high-Q cavity \cite{haro}, 
and of two internal states
of a $^{9}$Be$^{+}$ ion coupled with the vibrations in a trapping potential
\cite{wine}. In both cases one observes damped oscillations to a steady state
in which the population of each of the two levels approaches $1/2$.
A number of candidates have been already considered
as possible physical sources of decoherence in these cases. 
In the trapped ion
case of Ref.~\cite{wine}, fluctuations of classical parameters such 
as the intensity 
of the laser beams used to couple internal and vibrational degrees of 
freedom, or the voltage and the frequency of the trapping potentials have
been suggested. In
the case of the Rydberg atom in a high-Q cavity, dark counts of 
the atomic detectors, dephasing collisions with background gas or
stray magnetic fields within the cavity \cite{haro,pri}
have been proposed as possible sources of decoherence. Despite
this, a complete {\em quantitative} 
explanation of the observed decay rate of the Rabi oscillations
(see Ref.~\cite{nist}) in the two experiments is still
lacking. The only established fact is that,
differently from Ref.~\cite{prlha},
in both cases, decoherence has a {\em non-dissipative} origin. In fact,
the observed decay of the Rabi oscillations is much
faster than the energy relaxation rate in these experimental 
configurations. Moreover, the fact that in both cases
the population of each of the two levels asymptotically
approaches $1/2$ cannot be explained in terms of dissipative 
mechanisms as the photon leakage out of the cavity. 

A different approach to decoherence has been proposed in \cite{rb1}, 
where a {\em model-independent} formalism
has been derived to describe decoherence.
Here we shall adopt a more pragmatic point of view and we shall use
this formalism to explain in simple terms {\em both} Rabi oscillation
experiments, even
though they are realized in different physical situations.
The idea underlying the approach of Ref.~\cite{rb1}
is the fact
that the interaction time, i.e. the time interval in which
the effective Hamiltonian evolution takes place, is a random variable.
This randomness can have different origins depending on the 
studied system. For example, in the case of the Rydberg atom 
experiment \cite{haro}, the interaction time
is determined by the transit time of the velocity-selected atom
through the high-Q microwave cavity. 
This interaction time is random, due to 
fluctuations of the atomic velocities. 
This randomness implies having random phases $e^{-iE_{n}t/\hbar}$
in the energy eigenstates basis. The experimental results 
unavoidably average over these random phases and this 
leads to decoherence, i.e., to the decay of off-diagonal
matrix elements of the density operator in the energy basis.
Notice however that one would have the same phase fluctuations if
the Hamiltonian (and therefore the eigenvalues $E_{n}$) fluctuates instead
of the interaction time.
Therefore, as we shall see, our approach will give a 
generalized phase-destroying master equation,
able to describe many situations in which decoherence is associated
with random phases, originating for example from some 
frequency or interaction time fluctuations.

Let us consider an initial state $\rho(0)$ and consider the case
of a random evolution time. The experimentally observed state 
is not described by the usual density matrix of the whole system 
$\rho(t)$, but by its
time averaged counterpart \cite{rb1}
\begin{equation}
\bar{\rho}(t)= \int_{0}^{\infty} dt' P(t,t') \rho(t') \;,
\label{robar} 
\end{equation}
where $\rho(t') = \exp\{-iLt'\}\rho(0)$ is the usual unitarily evolved
density operator from the initial state and $L\ldots =[H,\ldots]/
\hbar$. Hence one can write
\begin{equation}
\bar{\rho}(t)= V(t) \rho(0) \;,
\label{robar2} 
\end{equation}
where $
V(t)= \int_{0}^{\infty} dt' P(t,t') e^{-iLt'}$.
In Ref.~\cite{rb1}, the function $P(t,t')$ has been 
determined so to satisfy the following conditions: i) $\bar{\rho}(t)$ must
be a density operator, i.e. it must be self-adjoint, positive-definite, and
with unit-trace. This leads to the condition that $P(t,t')$ must be
non-negative and normalized, i.e a probability density in $t'$ so that
Eq.~(\ref{robar}) is a completely positive mapping. ii) $V(t)$ satisfies
the semigroup property $V(t_{1}+t_2)=V(t_1)V(t_2)$, with $t_1,t_2 \geq 0$.
These requirements are satisfied by \cite{rb1}
\begin{eqnarray}
V(t) &=&\left(1+iL\tau\right)^{-t/\tau}
\label{fund} \\
P(t,t')&=& \frac{e^{-t'/\tau}}{\tau} \frac{(t'/\tau)^{(t/\tau)-1}}
{\Gamma(t/\tau)} \;,
\label{gamma}
\end{eqnarray}
so that $V(t)$ and $P(t,t')$ are connected by the so-called 
$\Gamma$-function integral identity
\cite{grad,note}.
The parameter $\tau$ characterizes the strength of the evolution time 
fluctuations.
When $\tau \rightarrow 0$, $P(t,t') \rightarrow \delta(t-t')$
so that $\bar{\rho}(t)=\rho(t)$ and $V(t)=\exp\{-iLt\}$ is the 
usual unitary evolution. However, for finite $\tau$, the evolution 
operator $V(t)$ of Eq.~(\ref{fund}) 
describes a decay of the off-diagonal matrix elements
in the energy representation, whereas the diagonal matrix elements remain
constant, i.e. the energy is still a constant of motion.
In fact, by differentiating with respect to time Eq.~(\ref{robar2})
and using (\ref{fund}),
one gets the following master equation for $\bar{\rho}(t)$
\begin{equation}
\dot{\bar{\rho}}(t)=-\frac{1}{\tau}\log\left(1+iL\tau\right) \bar{\rho}(t)
\label{ropun}
\end{equation}
If one expands the logarithm at second order in $\tau$, one obtains
\begin{equation}
\dot{\bar{\rho}}(t)=
-\frac{i}{\hbar}\left[H,\bar{\rho}(t)\right]-\frac{\tau}{\hbar^2}\left[
H,\left[H,\bar{\rho}(t)\right] \right] \;,
\label{ropun2}
\end{equation}
which is the well-known phase-destroying master equation \cite{milwal}.
Hence Eq.~(\ref{ropun}) appears as a {\em generalized} phase-destroying
master equation taking into account higher order terms in $\tau$.  
Notice, however, that the present approach is different from the
usual master equation approach in the sense that 
no perturbative and specific
statistical assumptions are made. 

We now apply this formalism to the two experiments
of Refs.~\cite{haro,wine}. In the experiment of Ref.~\cite{haro}, the 
resonant interaction between a quantized mode in a high-Q microwave 
cavity (with annihilation operator $a$)
and two circular Rydberg states ($|e \rangle $ and $|g \rangle $)
of a Rb atom is studied. This 
interaction is well described by the usual Jaynes-Cummings \cite{jc}
model, which in the interaction picture reads
\begin{equation}
H=\hbar \Omega_{R} \left(|e \rangle \langle g |a+|g \rangle \langle e |
a^{\dagger}\right) \;,
\label{jc}
\end{equation}
where $\Omega_{R}$ is the Rabi frequency.
The Rabi oscillations describing the exchange of excitations
between atom and cavity mode are studied by injecting
the velocity-selected Rydberg atom, prepared
in the excited state $|e \rangle $, in the high-Q cavity and measuring the
population of the lower atomic level $g$,
$P_{eg}(t)$ as a function of the interaction time $t$, which is varied 
by changing the Rydberg atom velocity. 
In the case of vacuum state 
induced Rabi oscillations, the decoherence effect is particularly
evident and the
Hamiltonian evolution according to Eq.~(\ref{jc}) predicts 
\begin{equation}
P_{eg}(t) =\frac{1}{2} \left(1- \cos \left(2\Omega_{R} 
t\right)\right)\;.
\label{rabi1}
\end{equation}
Experimentally instead, damped oscillations are observed, which are well
fitted by 
\begin{equation}
P_{eg}^{exp}(t) =\frac{1}{2} \left(1- e^{-\gamma t}
\cos \left(2\Omega_R t\right)\right) \;,
\label{fit}
\end{equation}
where the decay time fitting the experimental data
is $\gamma ^{-1} =40 \mu$sec \cite{pri} and the corresponding 
Rabi frequency is $\Omega_R/2\pi =25$ Khz. This decay of quantum 
coherence cannot be associated with photon leakage out of the cavity
because the cavity relaxation time is larger ($220$ $\mu$sec) 
and also because in this case one would have an asymptotic limit
$P_{eg}^{exp}(\infty)=1$. The damped behavior of Eq.~(\ref{fit}) is
instead easily obtained if one applies the approach described above.
In fact, from
the linearity of Eq.~(\ref{robar}), 
one has that the
time averaging procedure is also valid for mean values and matrix
elements of each subsystem. Therefore one has
\begin{equation}
\bar{P}_{eg}(t) = \int_{0}^{\infty} dt' P(t,t') P_{eg}(t') \;.
\label{pegbar} 
\end{equation}
Using Eqs.~(\ref{robar2}), (\ref{fund}), (\ref{gamma}) and (\ref{rabi1}),  
Eq.~(\ref{pegbar}) can be rewritten in the same form 
of Eq.~(\ref{fit})
\begin{equation}
\bar{P}_{eg}(t) =\frac{1}{2} \left(1- e^{-\gamma t}
\cos \left(\nu t\right)\right) \;,
\label{fit2}
\end{equation}
where 
\begin{eqnarray}
\label{gammalog}
\gamma &= &\frac{1}{2\tau} \log \left(1+4\Omega_{R}^2 \tau^2\right) \\
\nu &= &\frac{1}{\tau} {\rm arctg}\left(2\Omega_R \tau\right)
\end{eqnarray}
We note that in general the time averaging procedure
introduces not only a damping of the probability oscillations
but also a frequency shift. However, if the characteristic time $\tau$
is sufficiently small, i.e. $\Omega_R \tau \ll 1$, there is
no phase shift, $\nu \simeq 2\Omega_R$,
and 
\begin{equation}
\gamma = 2 \Omega_{R}^{2}\tau 
\label{gammapr}
\end{equation}
The fact that in Ref.~\cite{haro} the Rabi oscillation frequency essentially
coincides with the theoretically expected one, suggests that the time 
$\tau$ characterizing the fluctuations of the interaction time is 
sufficiently small so that it is reasonable to use Eq.~(\ref{gammapr}).
Using the above values for $\gamma$ and $\Omega_{R}$,
one can derive an estimate for $\tau$, so to
get $\tau \simeq 0.5$ $\mu$sec. This estimate is consistent with 
the assumption $\Omega_{R}\tau \ll 1$ we have made, but,
more importantly, it turns out to be
comparable to the experimental value of the uncertainty in the interaction time.
In fact, the fluctuations of the interaction time are mainly due to the 
experimental uncertainty
of the atomic velocity $v$. In fact, one has $t=\sqrt{\pi}w/v$, where $w$
is the cavity mode waist. Since $w=0.6$ cm, the mean velocity
is $\bar{v}\simeq 300$ m/sec and the velocity uncertainty is 
$\delta v/v =1\%$ (see Ref.~\cite{haro}), one has $\bar{t} =
\sqrt{\pi} w/\bar{v} \simeq 50$ $\mu$sec and $\tau \simeq \delta t
= \bar{t} \delta v/v =0.5$ $\mu$sec, 
which is just the estimate we have derived from the 
experimental values.
This simple argument supports the interpretation that
the decoherence observed in \cite{haro} is essentially
due to the randomness of 
the interaction time.

Let us now consider the case 
of the trapped ion experiment of Ref.~\cite{wine}, in which the 
interaction between two internal states ($|\uparrow \rangle
$ and $|\downarrow \rangle$) of a Be ion and the 
center-of-mass vibrations in the $z$ direction, induced by two
driving Raman lasers is studied. 
In the interaction picture with respect to the free vibrational and 
internal Hamiltonian, this interaction is described by the following
Hamiltonian \cite{nist}
\begin{eqnarray}
H &= &\hbar \Omega |\uparrow \rangle \langle \downarrow |
\exp\left\{i\left[\eta \left(
a e^{-i\omega_{z}t}+a^{\dagger}e^{i\omega_{z}t}\right)-\delta t+\phi
\right]\right\} \nonumber \\
&+ & H.C. \;,
\label{hgen}
\end{eqnarray}
where $a$ denotes the 
annihiliation operator for the vibrations along the $z$ direction, 
$\omega_{z}$ is the corresponding frequency and $\delta $ 
is the detuning between the 
internal transition and the frequency difference between the two Raman 
lasers. The Rabi frequency $\Omega$ is proportional to the 
two Raman laser intensities, and $\eta $ is the Lamb-Dicke parameter
\cite{wine,nist}.
When the two Raman lasers
are tuned to the first blue sideband, i.e. $\delta = 
\omega_{z}$, 
Hamiltonian (\ref{hgen}) predicts Rabi oscillations between $|\downarrow,n\rangle $
and $|\uparrow, n+1\rangle$ ($|n\rangle $ is a vibrational
Fock state) with a frequency \cite{nist}
\begin{equation}
\Omega_n = \Omega \frac{e^{-\eta^2/2}}{\sqrt{n+1}}\eta L_{n}^{1}(\eta ^2) \;,
\label{omn}
\end{equation}
where $L_{n}^{1}$ is the generalized Laguerre polynomial.
These Rabi oscillations have been experimentally verified
by preparing the initial state $|\downarrow , n
\rangle $, (with 
$n$ ranging from $0 $ to $16$) and measuring the probability
$P_{\downarrow}(t)$ as a function of the interaction time $t$, which is varied
by changing the duration of the Raman laser pulses. 
Again, as in the cavity QED experiment of \cite{haro}, 
the experimental Rabi oscillations are damped and well fitted by \cite{wine,nist}
\begin{equation}
P_{\downarrow}(n,t) =\frac{1}{2} \left(1+ e^{-\gamma_n t}
\cos \left(2\Omega_n t\right)\right) \;,
\label{fitwine}
\end{equation}
where the 
measured oscillation frequencies $\Omega_n$ are in very good agreement with 
the theoretical prediction (\ref{omn})
corresponding to the measured Lamb-Dicke parameter $\eta =0.202$ \cite{wine}.
As concerns the decay rates $\gamma_n$, 
the experimental values are fitted in \cite{wine} by
\begin{equation}
\gamma_n = \gamma_0 (n+1)^{0.7}
\label{o7}
\end{equation}  
where $\gamma_0=11.9$ Khz. 
This power-law scaling has been investigated in Refs.~\cite{milb,murao}, but
a clear explanation of this behavior of the decay rates is still lacking.
On the contrary, the scaling law (\ref{o7}) can be accounted for
in the previous formalism if we consider the small $\tau$ limit
of Eq.~(\ref{gammapr}), which is again suggested by the
fact that the experimental and theoretical predictions for the
frequencies $\Omega_n$ agree. 
In fact, the $n$-dependence of the theoretical prediction
of Eq.~(\ref{omn}) for $\eta =0.202$ is well approximated, within 10 \%,
by the power law dependence 
\begin{equation}
\Omega_n \simeq \Omega_0 (n+1)^{0.35} \;,
\label{o35}
\end{equation}  
so that, using Eq.~(\ref{gammapr}) with $\Omega_{R}$ replaced by
$\Omega_{n}$, one has immediately the power law dependence
$(n+1)^{0.7}$ of Eq.~(\ref{o7}).
The value of the parameter $\tau$ can be obtained by matching the
values corresponding to $n=0$, and using Eq.~(\ref{gammapr}), that is
$\tau = \gamma_0/2\Omega_0 ^2 \simeq 1.5 \cdot 10^{-8}$ sec,
where we have used the experimental value $\Omega_{0}/2\pi = 94$ Khz.
However, this value of the parameter $\tau$ cannot be explained in terms
of some interaction time uncertainty, such as the time jitter 
of the Raman laser pulses, which is experimentally found to be
much smaller \cite{pri2}. In this case, instead, the observed decoherence
can be attributed, as already suggested in \cite{nist,milb,murao}, to the 
fluctuation of the Raman laser
intensities, yielding a fluctuating Rabi frequency parameter
$\Omega (t)$ of the Hamiltonian (\ref{hgen}). In this case
the evolution is driven by a fluctuating Hamiltonian $H(t)=
\hbar \Omega(t) \tilde{H}$, where $\tilde{H}=H/\Omega$ in Eq.~(\ref{hgen}), 
so that
\begin{equation}
\rho(t) = \exp\left\{-i \tilde{L}\int_{0}^{t}d\xi \Omega(\xi)\right\}\rho(0)
=e^{-i\tilde{L} A(t)} \rho(0)
\label{rotild}
\end{equation}
where $\tilde{L}=[\tilde{H},\ldots ]/\hbar$ and we have defined
the positive dimensionless random variable $A(t)=\int_0 ^t d\xi \Omega(\xi)$, 
which is proportional to the pulse area. It is now easy to understand
that the physical situation is analogous to that characterized by
a random interaction time considered above, with $L$ replaced
by $\tilde{L}$ and $t'$ by $A(t)$. One has again phase
fluctuations in the energy basis representation and, in analogy with
Eq.~(\ref{robar}), one considers 
an averaged density matrix
\begin{equation}
\bar{\rho}(t)= \int_{0}^{\infty} dA P(t,A) e^{-i\tilde{L} A} \rho(0) \;.
\label{robara} 
\end{equation}
Imposing again that $\bar{\rho}(t) $
must be a density operator and the semigroup property, one finds 
results analogous to Eqs.~(\ref{fund}) and (\ref{gamma})
\begin{eqnarray}
V(t) &=&\left(1+i\tilde{L}\Omega\tau\right)^{-t/\tau}
\label{funda} \\
P(t,A)&=& \frac{e^{-A/\Omega\tau}}{\Omega\tau} 
\frac{(A/\Omega\tau)^{(t/\tau)-1}}
{\Gamma(t/\tau)} \;,
\label{agamma}
\end{eqnarray}
where, the parameter $\Omega$ of Eq.~(\ref{hgen})
plays now the role of a 
mean Rabi frequency. In fact, consistently with the probability distribution
of Eq.~(\ref{agamma}), one has $\Omega =\langle A \rangle /t$.
The scaling time $\tau$ characterizes in this case the strength
of the pulse area fluctuations, since from Eq.~(\ref{agamma}), 
one has $\sigma^2(A)=
\langle A^2\rangle - \langle A \rangle ^2 = \Omega^2 t\tau $.
The estimated value of $\tau$ is reasonable 
since it corresponds to a fractional error of the pulse area
$\sqrt{\sigma^2(A)}/\langle A \rangle = \sqrt{\tau/t}$ of $10 \%$ 
for a pulse duration of $t=1$ $\mu$sec, and which is decreasing for
increasing pulse durations.

The present analysis shows many similarities with that of Ref.~\cite{milb}
which also tries to explain the decay of the Rabi oscillations in the ion trap
experiments of \cite{wine} in terms of laser intensity fluctuations.
The authors of Ref.~\cite{milb} in fact use a phase 
destroying master equation coinciding with the second-order expansion
(\ref{ropun2}) of our generalized master equation of Eq.~(\ref{ropun})
(see Eq.~(16) of Ref.~\cite{milb} with the identifications $G \leftrightarrow
H/\hbar$ and $\Gamma \leftrightarrow \tau$) and moreover derive the
same numerical estimate for the pulse area fluctuation strength 
$\Gamma \leftrightarrow \tau$. Despite this similarities, 
they do not recover the scaling (\ref{o7}) of the decay rates 
$\gamma_n$ only because they
do not use the general expression of the Rabi frequency (\ref{omn}),
(and which is well approximated by the power law (\ref{o35}))
but its Lamb-Dicke limit
$ \Omega_n = \Omega_0 (n+1)^{0.5} $,
which is valid only when $\eta \ll 1$.
There is however another, more fundamental, difference between 
our approach and that of Ref.~\cite{milb}. They
assume from the beginning that the laser intensity fluctuations have
a white and gaussian character, while we make no {\em a priori}
assumption on the statistical properties of the pulse area $A$.
We derive these properties, i.e. the probability distribution (\ref{agamma}),
only from the very general semigroup condition, and it is interesting to note
that this condition yields a gaussian probability distribution
for the pulse area only as a limiting case. In fact, from Eq.~(\ref{agamma})
one can see that $P(t,A)$ tends to become a gaussian with the same
mean value
$\Omega t$ and the same variance $\Omega^2 \tau t$ only 
in the large time limit
$t/\tau \gg 1$. This non-gaussian character of $P(t,A)$
can be traced back to the fact that $P(t,A)$ must be definite
and normalized in the interval $0\leq A<+\infty$ and not in
$-\infty < A < +\infty$. Notice that at $t=\tau$, Eq.~(\ref{agamma})
assumes the exponential form $P(t=\tau,A)=e^{-A/\Omega \tau}/\Omega \tau$.
 Only at 
large times $t$ the random variable $A$ becomes the sum of many independent
contributions and assumes the gaussian form. 
Due to the non-gaussian nature of the random variable $A$, we find that
the more generally valid phase-destroying master equation is given by
Eq.~(\ref{ropun}) (with $L$ replaced by $\Omega\tilde{L}$), 
whose predictions significantly depart from its
second order expansion (\ref{ropun2}) (corresponding
to the gaussian limit) as soon as $\tau$
becomes comparable with the typical timescale of the system under study, 
which, in the present case, is the inverse of the Rabi frequency.

In conclusion, we have presented a model-independent theory
for non-dissipative decoherence, 
able to provide a simple and {\em unified} description of the 
same decoherence phenomenon
observed in two Rabi oscillations experiments
which were performed under different situations.
A simple way to test experimentally our prediction
is to check that the dependence of the decay rate as a function
of the Rabi frequency is given by Eq.~(\ref{gammalog}). 
One should observe a transition from a quadratic
dependence to a logarithmic dependence, increasing the value
of the Rabi frequency, or of $\tau$.

Discussions with J.M. Raimond and D. Wineland are greatly
acknowledged. This work has been supported by
MURST under the ``Cofinanziamento''.

\bibliographystyle{unsrt}

\begin{thebibliography}{99}
\bibitem{note}In Ref.~\protect\cite{rb1} a more general expression for $P(t,t')$
and $V(t)$, depending on two parameters $\tau_1$ and $\tau_2$ is derived.
We choose $\tau_1=\tau_2 = \tau$ because in the experiments considered 
here, the effective interaction time $\langle t'\rangle =t \tau_1/\tau_2$
\protect\cite{rb1}
has to coincide with the ``laboratory time'' $t$, implying 
therefore $\tau_1=\tau_2$.
\end{thebibliography}

\end{document}